\documentclass[onecolumn,aps,nofootinbib,notitlepage]{revtex4-1}
\usepackage[margin=2cm]{geometry}
\usepackage{amsfonts}
\usepackage{amsmath}
\usepackage{amsthm} 
\usepackage{amssymb}
\usepackage[svgnames]{xcolor}
\usepackage[hidelinks]{hyperref}
\hypersetup{
    colorlinks=true,
    linkcolor=NavyBlue,
    filecolor=NavyBlue,      
    urlcolor=NavyBlue,
    citecolor=NavyBlue,
    }
 
\usepackage{graphics}
\usepackage{graphicx}  
\usepackage{caption}
\usepackage{subcaption} 
\usepackage{adjustbox}
\usepackage{gensymb}
\usepackage{breqn}
\usepackage{braket}
\usepackage{gensymb} 
\usepackage{slashed}
\usepackage[compat=1.0.0]{tikz-feynman}

\begin{document} 
\begin{center}
\hfill  MI-TH-1929
\end{center}

\title{\large{\textbf{ $(g-2)_{\mu,e}$ and the ANITA anomalous events in a three-loop neutrino mass model}}}
\author{\normalsize{Mohammad Abdullah$^{\bf 1}$\footnote{mabdullah@tamu.edu}, Bhaskar Dutta$^{\bf 1}$\footnote{dutta@physics.tamu.edu}, Sumit Ghosh$^{\bf 1}$\footnote{ghosh@tamu.edu},  Tianjun Li$^{\bf {2,3}}$\footnote{tli@itp.ac.cn}} \\
\vspace{1.0cm}
\normalsize\emph{$^{\bf 1}$Mitchell Institute for Fundamental Physics and Astronomy, Department of Physics  and Astronomy, Texas A$\&$M University, 
College Station, Texas 77843, USA}\\
\normalsize \emph{$^{\bf 2}$CAS Key Laboratory of Theoretical Physics, Institute of Theoretical Physics, \protect \\ Chinese Academy of Sciences, Beijing, 100190, People's Republic China}\\
\normalsize\emph{$^{\bf 3}$School of Physical Sciences, University of Chinese Academy of Sciences, \protect \\ Beijing 100049, People's Republic China}\\
\vspace{1.0cm} 
}
 
\begin{abstract} 
The most recent measurement of the fine structure constant leads to a 2.4 $\sigma$ deviation in the electron anomalous magnetic moment 
$g_{e}$-2, while the muon anomalous magnetic moment $g_\mu$-2 has a long standing 3.7 $\sigma$ deviation in the opposite direction. We show that 
these deviations can be explained in a three-loop neutrino mass model based on an $E_6$ Grand Unified Theory. We also study the impact 
such a model can have on the anomalous events observed by the ANITA experiment and find an insufficient enhancement of the event rate.
\end{abstract}
\maketitle


\section{Introduction}

The Standard Model~(SM) of particle physics, while a very successful and mathematically consistent theory, is challenged by a variety of experimental and theoretical puzzles. The neutrino oscillation data~\cite{Fukuda:1998mi,Ahmad:2002jz} conclusively require neutrinos to have small but non-zero masses, and although a plethora of models exist in the literature for generating such masses, keeping the new dynamics at an energy scale accessible by current and near-future experiments proves nontrivial. 

There are, additionally, some less certain but equally intriguing puzzles. One of the long-standing deviations of the experimental data from the theoretical predictions of the SM is the anomalous magnetic moment of the muon, $a_\mu=(g_\mu-2)$. There is a 3.7~$\sigma$ discrepancy between the experimental results~\cite{Bennett:2006fi,Tanabashi:2018oca} and theoretical predictions~\cite{Davier:2017zfy,Blum:2018mom,Keshavarzi:2018mgv,Davier:2019can}. This has recently been compounded with a 2.4~$\sigma$ discrepancy between the experiment~\cite{Hanneke:2008tm, Hanneke:2010au} and theory~\cite{Aoyama:2017uqe} values of $a_e$. Moreover, the deviations are in opposite directions, meaning that the most straightforward Beyond the SM (BSM) explanations are insufficient. Also, $\Delta a_e/\Delta a_\mu$ does not follow the lepton mass scaling $m_e^2/m_\mu^2\sim 2.25\times 10^{-5}$, which means that a model with new flavor structure in the leptonic sector would be required to explain the discrepancy.

Another interesting experimental result is the observation of the two up-going ultra-high energy cosmic ray air shower events by the ANITA Collaboration (ANtarctic Impulsive Transient Antenna) \cite{Gorham:2016zah,Gorham:2018ydl}. In principle, the events could be explained by a $\nu_\tau$ that up-scatters inside the Earth into a $\tau$ and then decays hadronically upon emergence. However, the Earth is far too opaque for both the neutrino and the charged lepton at the measured energy (about 0.6 EeV) and angle (about 30\degree \;below the horizon) \cite{Fox:2018syq}, requiring a neutrino flux that exceeds the current limits from the Pierre Auger Observatory \cite{Aab:2015kma} and IceCube \cite{Aartsen:2018vtx}. 

From a big picture perspective, the unification of the electromagnetic and weak forces, the cancellation of gauge anomalies, and the near intersection of the gauge couplings at high energies in the SM all hint at a Grand Unified Theory (GUT). In a previous work, an $E_6$ GUT inspired model was shown to accommodate the neutrino masses and mixing parameters with TeV scale physics \cite{Dutta:2018qei}. Such a low scale was made possible by forbidding mass diagrams below 3-loop order. The flavor structure introduced in that work has the potential to explain the $g-2$ anomalies. In this work, we show that the scalar and fermionic degrees of freedom in that model together with the correct flavor structure  can provide the necessary corrections to $g-2$ of both the muon and electron. We also study the implications on the ANITA observations and find that, while the expected event rate can be enhanced, explaining the two anomalous events remains challenging.

The outline of the paper is as follows: in Sec.~\ref{Model}, we present the model. We define all the necessary physical fermions and scalar particles in Sec.~\ref{physical}. In Sec.~\ref{neutrino mass}, we briefly discuss the neutrino mass generation. We study the $g-2$ for both electron and muon in Sec.~\ref{sec:g-2}. In Sec.~\ref{Anita}, we discuss the ANITA anomalous events. Finally, we conclude in Sec.~\ref{Conclusion}.


\section{The $SU(3)_C\times SU(2)_L \times SU(2)_N \times U(1)_Y$ Model from $E_6$ GUT} \label{Model}

In this section, we briefly describe the gauge symmetry and the field content of the model introduced in~\cite{Dutta:2018qei}. A maximal subgroup of $E_6$ is $SU(6) \times SU(2)_N$, while a maximal subgroup of $SU(6)$ is $SU(5) \times U(1)'$. We assume the latter is broken down to generate the SM gauge groups while the $SU(2)_N$ survives down to low energy and assumed not to affect the electric charge operator which retains the form $Q=T_{3L}+Y$. The particle content under $SU(3)_C\times SU(2)_L \times SU(2)_N \times U(1)_Y$ is as follows

{\small
 \[  Q_i\sim  \left( \begin{array}{c} u_i \\ d_i \end{array} \right) \sim (3,2,1,\frac{1}{6}) ,\ \ \ \ \  U_i^c \sim (\bar{3},1,1,-\frac{2}{3}),\ \ \ \ \  D_i^c\sim  \left( {d^\prime}_i^c \ \   d_i^c \right) \sim (\bar{3},1,2,\frac{1}{3})~,~\, \] \[  D_i\sim (3,1,1,-\frac{1}{3}),\ \ \ \ \ L_i\sim  \left( \begin{array}{cc} E_i^0 &\nu_{i} \\ E_i^- & e_i^- \end{array} \right) \sim (1,2,2,-\frac{1}{2}) , 
\ \ \ \ \  E_i^c \sim (1,1,1,1)~,~\, \]  \[  {L_i}^\prime \sim   \left( \begin{array}{c} E_i^+ \\ \bar{E_i}^{0} \end{array} \right) \sim  (1,2,1,\frac{1}{2}),\ \ \ \ \ N_i^c\sim \left( n_{1i}^c \ \ n_{2i}^c \right)\sim (1,1,2,0)~,~\,\] \[  F_i \sim  \left( \begin{array}{cc} F_{3i} & F_{1i} \\ F_{2i} & -F_{3i} \end{array} \right) \sim  (1,1,3,-1), \ \ \ \ \  {F_i}^c \sim  \left( \begin{array}{cc} {F^c_{3i}} & {F^c_{1i}} \\ {F^c_{2i}} & -{F^c_{3i}} \end{array} \right) \sim(1,1,3,1)~,~\, \] }  
where $i$  $=$ $1,2,3$, are the generation index. The fundamental representation of $E_6$ i.e. $\bf{27}$ can accommodate $Q_i$, $U_i^c$, $D_i^c$, $D_i$, $L_i$, $E_i^c$, $L'_i$, and $N_i^c$, while the vector-like fermions $F_i$ and $F_i^c$ come from the \textbf{351} and  \textbf{$\overline{\textrm{351}}$} representations of $E_6$, respectively. The embedding of all fields in full GUT multiplets ensures that the gauge anomalies are automatically cancelled.

The scalar sector of the model has four scalar field with the following charge assignment
{\small  \[H_d  \sim  (1,2,2,-\frac{1}{2}),~~~ H_u  \sim  (1,2,1,\frac{1}{2})~,~\,~~~ S^0  \sim (1,1,2,0),\ \ \ \ \  T  \sim  (1,2,2,\frac{3}{2})~.~\, \]  }
One $\bf{27}$ representation  can give  $H_d$, $H_u$, and $S^0$ while one \textbf{650} gives the bi-doublet scalar field $T$.  

The most general renormalizable scalar potential is the following \footnote{We suppress the gauge indices and refer the reader to the previous work where this model was introduced \cite{Dutta:2018qei}.}
\begin{dmath} \label{potential} 
 V =m_1^2\; \text{Tr}[{H_d}^{\dag}\cdot{H_d}]
+m_2^2\; {H_u}^{\dag}\cdot{H_u}
+m_s^2\;S^0\cdot{S^0}^{\dag}
+m_T^2\; \text{Tr}[{T}^{\dag}\cdot{T}]
+\frac{\lambda_2}{2}({H_u}^{\dag}\cdot{H_u})^2
+\frac{\lambda_1}{2}\; \left(\text{Tr}[{H_d}^{\dag}\cdot{H_d}]\right)^2
+\frac{\lambda_3}{2}\; \text{Tr}[{H_d}^{\dag}\cdot{H_d}\cdot{H_d}^{\dag}\cdot{H_d}] 
+\frac{\lambda_s}{2}(S^0\cdot{S^0}^{\dag})^2
+\frac{\lambda_6}{2}\; \left(\text{Tr}[{T}^{\dag}\cdot{T}]\right)^2
+\frac{\lambda_7}{2}\; \text{Tr}[{T}^{\dag}\cdot{T}\cdot{T}^{\dag}\cdot{T}]
+\lambda_4({H_u}^{\dag}\cdot{H_u})\; \text{Tr}[{H_d}^{\dag}\cdot{H_d}]
+\lambda_5{H_u}^{\dag}\cdot{H_d}\cdot{H_d}^{\dag}\cdot{H_u}
+\lambda_8({H_u}^{\dag}\cdot{H_u})(S^0\cdot{S^0}^{\dag})
+\lambda_9(S^0\cdot {S^0}^{\dag})\; \text{Tr}[{H_d}^{\dag}{H_d}]
+\lambda_{10} S^0\cdot{H_d}^{\dag}\cdot{H_d}\cdot {S^0}^{\dag}_\gamma
+\lambda_{11}(S^0\cdot {S^0}^{\dag})\; \text{Tr}[{T}^{\dag}{T}]
+\lambda_{12}  S^0\cdot{T}^{\dag}\cdot{T}\cdot {S^0}^{\dag}
+\lambda_{13} ({H_u}^{\dag}\cdot{H_u})\; \text{Tr}[{T}^{\dag}\cdot{T}]
+\lambda_{14} {H_u}^{\dag}\cdot{T}\cdot{T}^{\dag}\cdot{H_u}_{\gamma}
+\lambda_{15} \; \text{Tr}[{H_d}^{\dag}\cdot{H_d}]\; \text{Tr}[{T}^{\dag}\cdot{T}]
+\lambda_{16} \; \text{Tr}[{H_d}^{\dag}\cdot{H_d}\cdot{T}^{\dag}\cdot{T}]
+\lambda_{17}  \; \text{Tr}[{H_d}^{\dag}\cdot{T}]\; \text{Tr}[{T}^{\dag}\cdot{H_d}]
+\lambda^{\prime}[\widetilde{H}_u\cdot{H_d}\cdot \widetilde{S}^0+H.c.] 
+\lambda \left[\; \text{Tr}[\widetilde{T}^T\cdot{\widetilde{H}_d}]\; \text{Tr}[\widetilde{H}^T_d\cdot{\widetilde{H}_d}]+\widetilde{T}^T\cdot{\widetilde{H}_d}\cdot \widetilde{H}^T_d \cdot{\widetilde{H}_d}+H.c.\right]~,~\, \end{dmath} 
where all the parameters are real and $\widetilde{X}$ denotes a single contraction with an antisymmetric tensor.

The Yukawa potential is given by: 
\begin{dmath} 
\label{eq:yukawa}
-\mathcal{L}_{Yukawa} = 
y_{1ij}\; \text{Tr}[\widetilde{L}_{i}\cdot \widetilde{F}_{j}\cdot \widetilde{T}^T]
+ y_{2ij} \; \text{Tr}[\widetilde{L}_{i}\cdot \widetilde{F}^{\dag}_{j}\cdot {\widetilde{H}_d}]
+y_{3ij}\widetilde{Q}_{i}\cdot{H_d}\cdot \widetilde{D}^c_{j}
+y_{4ij}\widetilde{Q}_{i}\cdot{H_u}U^c_{j}\\
+y_{5ij}\widetilde{D}^c_{i}\cdot {S^0}D_j
+y_{6ij}{\widetilde{L}^\prime}_{j}\cdot {L_i} \cdot {\widetilde{S}^0}
+y_{7ij}{\widetilde{L}^\prime}_{i} \cdot {H_d} \cdot {\widetilde{N}^c}_{j}
+y_{8ij} \; \text{Tr}[\widetilde{L}^T_{i} \cdot {\widetilde{H}_d}]E_j^c
+y_{9ij}\; \text{Tr}[\widetilde{H}_{u} \cdot  L_{i} \cdot \widetilde{N}_{j}^c]
+{ \small \frac{1}{2}M_{ij} {F}_i F^c_j} + m_{Nij}N_i^cN_j^c ~,~\,
\end{dmath}

 For simplicity, we assume $M_{ij}= M_i \delta_{ij}$. We impose a discrete $Z_2$ symmetry such that only $N_i^c$ is odd under the $Z_2$ symmetry while all the other particles are even. One implication of this $Z_2$ symmetry is that the $y_{7ij}$ and  $y_{9ij}$  terms in Eq. \ref{eq:yukawa} are forbidden. The lightest component of $N_i^c$ is therefore stable and may contribute to the dark matter density.


\section{The physical scalars and fermions} \label{physical} 

We define the physical scalar particles and fermions in the mass basis necessary for the calculations in this section. The total number of scalar degrees of freedom is 24, out of which 6 are eaten by the massive gauge bosons. This leaves us with 18 physical scalars. After the spontaneous symmetry breaking the Higgs scalars acquire vacuum expectation values (vevs) and we can write them as
\begin{eqnarray} \label{scalar_defs1}
H_d  &\sim& \left( \begin{array}{cc} \frac{1}{\sqrt{2}}(v_1+\rho_1+i\eta_1) &  \frac{1}{\sqrt{2}}(\rho_3+i\eta_3)  \\
 {\phi_1}^- & {\phi_3}^- \end{array} \right) ~,~\,  
 T \sim  \left( \begin{array}{cc} {T_1}^{++} &  {T_2}^{++} \\ {T_1}^+ &  {T_2}^+ \end{array} \right) ~,~\,  \nonumber\\ 
H_u  &\sim&   \left( \begin{array}{c}  {\phi_2}^+  \\  \frac{1}{\sqrt{2}}(v_2+\rho_2+i\eta_2)  \end{array} \right) ~,~\ \ \  
S^0 \sim \left(   \frac{1}{\sqrt{2}}(\rho_{1s}+i\eta_{1s}) \ \ \ \ \   \frac{1}{\sqrt{2}}(v_s+\rho_{2s}+i\eta_{2s}) \right) ~,~ 
\end{eqnarray}
where the vev of $S^0$ ($v_s$) is $\mathcal{O}$(10-100) TeV and breaks the $SU(2)_N$ symmetry, while the vevs of
$H_d$ and $H_u$ ($v_1$ and $v_2$ respectively) are $\mathcal{O}$(10-100) GeV and break the electroweak gauge symmetry. 

The charged states $\phi_1^\pm$ and $\phi_2^\pm$ will mix with a mixing angle $\beta$ and give 
the charged physical scalars  $h_1^\pm$ and $h_2^\pm$ 
respectively with masses $m_{h_1}$ and ${m_{h_2}}$ 
\begin{align} h_{1}^\pm =& \cos \beta\; \phi_1^\pm +\sin\beta\; \phi_2^\pm ~,~\,\nonumber \\  
h_{2}^\pm =& -\sin\beta\; \phi_1^\pm +\cos\beta \;\phi_2^\pm~,~\,\end{align}
where $\text{tan}\beta=v_2/v_1$. Similarly, four more charged physical scalars $H_1^\pm$ and $H_2^\pm$ with masses $m_{H_1}$ and ${m_{H_2}}$ respectively can arise from the mixing of $\phi_3^\pm$ and $T_2^\pm$
\begin{align} H_{1}^\pm =& \cos \theta \;\phi_3^\pm +\sin\theta\; T_2^\pm ~,~\,
\nonumber \\  H_{2}^\pm =& -\sin\theta \;\phi_3^\pm +\cos\theta\; T_2^\pm~.~\,\end{align}

There is a total of five neutral CP even states, out of which $\rho_1$, $\rho_2$ and $\rho_{2s}$ will mix resulting in three physical neutral scalars $s_1$, $s_2$ and $s_{2s}$ with masses $m_{s_1}$, $m_{s_2}$, and $m_{s_{2s}}$, respectively. We identify the $s_1$ as the SM physical Higgs field, $h$ with mass $m_h$. The fields in the mass basis, $s_i$ are related to those in the interaction basis, $\rho_i$ by a rotation matrix $R$ which can be parametrized with three angle $\theta_1,\theta_2 ~\mbox{and} ~\theta_3$ as follows
\begin{equation} R^{-1} = \left( \begin{array}{ccc} c_{\theta_1} c_{\theta_3}-c_{\theta_2}s_{\theta_1}s_{\theta_3} & -c_{\theta_1}s_{\theta_3}-c_{\theta_2}c_{\theta_3}s_{\theta_1} & s_{\theta_1}s_{\theta_2} \\ c_{\theta_3}s_{\theta_1}+c_{\theta_1}c_{\theta_2}s_{\theta_3} & c_{\theta_1}c_{\theta_2}c_{\theta_3}-s_{\theta_1}s_{\theta_3} & -c_{\theta_1}s_{\theta_2} \\ s_{\theta_2}s_{\theta_3} & c_{\theta_3}s_{\theta_2} & c_{\theta_2} \end{array} \right) \equiv\left( \begin{array}{ccc} c_{11} & c_{12} & c_{13} \\ c_{21} & c_{22} & c_{23} \\ c_{31} & c_{32} & c_{33} \end{array} \right) ~,~\, \end{equation}
where $\rho = R^{-1} s $, $c_{\theta_i}=\cos \theta_i$ and $s_{\theta_i}=\sin \theta_i$ ($i=1,2,3$). We can parametrize the mixing of three neutral CP-odd states, $\eta_1$, $\eta_2$, and $\eta_{2s}$ in a similar way with $\eta = R^{\prime^{-1}} s^\prime $, and the three angles are denoted as $\theta^\prime_1,\theta^\prime_2 ~\mbox{and} ~\theta^\prime_3$. We only get one physical neutral pseudoscalar $s^\prime_1$ with mass $m_{s^\prime_1}$. The interaction states can be written as, $\eta_1=c^\prime_{13} s^\prime_1$, $\eta_2=c^\prime_{23} s^\prime_1$, and $\eta_{2s}=c^\prime_{33} s^\prime_1$.

The rest of the two neutral CP-even states $\rho_3$ and $\rho_{1s}$ give one physical neutral scalar $s_3$ with mass $m_{s_3}$. The $2\times 2$ mixing matrix can be parametrize by an angle $\psi$ as
\begin{equation} R_{2\times 2}(\psi) =  \left( \begin{array}{cc} \cos \psi &  \sin \psi \\ -\sin \psi &  \cos \psi \end{array} \right)~.~ \end{equation} The interaction states can be written as $\rho_3  = -\sin \psi s_3$ and $\rho_{1s}  = \cos \psi s_3$. Similar mixing happens between the two neutral CP-odd states $\eta_3$ and $\eta_{1s}$. We parametrize the mixing matrix with an angle $\psi^\prime$. One physical pseudoscalar $s^\prime_3$ with mass $m_{s^\prime_3}$ is generated. In terms of this physical pseudoscalar, we write $\eta_3  = -\sin \psi^\prime s^\prime_3$ and $\eta_{1s}  = \cos \psi^\prime s^\prime_3$.

For the purpose of calculating the contributions to the anomalous magnetic moment and the ANITA observations, the relevant terms from Eq.~\ref{eq:yukawa} can be written as follows
\begin{dmath} \label{eq:g-22} 
-\mathcal{L} =  y_{2ii}(- E_i^0 \phi_1^-+\nu_{i} \phi_3^-+E_i^- \phi_1^0 -  e_i^-\phi_3^0 )F_{3i}^c +y_{3ij}[(u_i\phi_1^--d_i\phi_1^0)d^c_j-(u_i\phi_3^--d_i\phi_3^0) {d^\prime}_j^c]+ y_{5ij}( {d^\prime}_i^c S_2^0-d_i^cS_1^0)D_j +y_{6ii}[(E_i^0\bar{E}_i^0-E_i^-E_i^+)S_2^0-(\nu_i\bar{E}_i^0-e_i^-E_i^+)S_1^0]+ y_{8ii}(E_i^0\phi_3^--\nu_i\phi_1^--E_i^-\phi_3^0+e_i^-\phi_1^0)E_i^c+{ \small \frac{1}{2}M_ii {F}_i F^c_i} +{\rm H.C.}~,~\,
\end{dmath}
where $\phi^0_{1,2,3,}$ are neutral components of $H_u$ and $H_d$ (see Eq.~\ref{scalar_defs1}). The vector-like leptons $E^-_i$ and $F_{3i}$  mix and give rise to two charged physical vector-like leptons given by, 
\begin{align} f_{1i}^+ =& \cos{\theta_f} E_i^+ -\sin{\theta_f} F_{3i}^c \nonumber \\  f_{2i}^+ =& \sin{\theta_f} E_i^+ +\cos{\theta_f} F_{3i}^c\end{align}  respectively with masses $m_{f_{1i}}$ and $m_{f_{2i}}$, where the mixing angle,  $\theta_f$, can be determined by diagonalizing the mass matrix. We summarize all the relevant fields in Table~\ref{table:particle}.

\begin{table}[h]
\footnotesize
\centering
\begin{tabular}{ |c|c|c|c|c|}
\hline
\hline
Particle type	&	Particles 	&	 Mass parameters	&	 Mass values	&	Possible final states at LHC\\\hline

Charged scalars	&	$h_1^\pm,~h_2^\pm$ 	&	 $m_{h_1},~m_{h_2}$  	&	 $m_{h_{1,2}}\sim \mathcal{O}(500)$~GeV 	&\begin{tabular}{@{}c@{}}$h_{1,2}^\pm \rightarrow u_i\bar{d_j}+{E^0_i}+{E^0_j}$, \\ $ u_i\bar{d_j}$ \end{tabular}  \\\cline{2-5}

&$H_1^\pm,~H_2^\pm$&$m_{H_{1}},~m_{H_2}$&\begin{tabular}{@{}c@{}}$m_{H_{1}}\sim \mathcal{O}(500)$~GeV \\ $m_{H_{1}}\sim \mathcal{O}(5)$~TeV \end{tabular}&\begin{tabular}{@{}c@{}}$ H_{1,2}^\pm \rightarrow u_i\bar{d_j}+{\nu_i}+{E^0_j}$, \\ $d_i\bar{d_j}+{e_j^-}+{E^0_i}$ \end{tabular}\\ \hline
Neutral scalars& \begin{tabular}{@{}c@{}}$h,~s_2$ \\ $s_{2s},~s_3$ \end{tabular}&\begin{tabular}{@{}c@{}}$m_{h},~m_{s_2}$ \\ $m_{s_{2s}},~m_{s_3}$ \end{tabular}&\begin{tabular}{@{}c@{}}$m_{h}=125$ GeV \\ $m_{s} \sim \mathcal{O}(500)$ GeV  \end{tabular}&\begin{tabular}{@{}c@{}}$s_2,s_{2_s} \rightarrow e_i^+e_i^-$, \\ $d_i\bar{d_i}$ \end{tabular} \\ \hline
Neutral pseudoscalar&$s^\prime_1,s^\prime_3$&$m_{s^\prime_1},~m_{s^\prime_3}$&$m_{s^\prime} \sim \mathcal{O}$(500) GeV&\begin{tabular}{@{}c@{}}$s^\prime_1,s^\prime_3 \rightarrow e_i^+e_i^-$, \\ $d_i\bar{d_i}$ \end{tabular} \\\hline
Charged vector-like leptons&$f^\pm_1,~f^\pm_2$&$m_{f_1},~m_{f_2}$&$m_f \sim \mathcal{O}(100)$ GeV &\begin{tabular}{@{}c@{}}$f_{1,2i}\rightarrow u_i\bar{d_j}+E_i^0$ \\ $d_i\bar{d_j}+{e_i^+}+\bar{E^0_i}+\bar\nu_i$ \end{tabular}\\\hline
Neutral vector-like leptons&$\bar{E}^0,~E^0$&$m_{\bar{E}^0},~m_{E^0}$&$m_{\bar{E}^0}> m_{E^0}\sim\mathcal{O}(100)$GeV&\begin{tabular}{@{}c@{}}$\bar{E}^0_i \rightarrow e_i^+e_i^- +E^0_i$ \\ $E^0_i \rightarrow \nu_i + n_{1i} \bar{n}_{2j}$ \end{tabular}\\\hline
New gauge bosons&$X_1^\mu,~X_2^\mu,~X_3^\mu$&$m_{X_1},~m_{X_2},~m_{X_3}$&$m_X \ge 3.6$ TeV&\begin{tabular}{@{}c@{}}$X_3^\mu \rightarrow e_i^+e_i^-$, \\ $d_i\bar{d_i}$ \end{tabular} \\\hline
Charged vector-like quark& $D$& $m_D$& $m_D \sim \mathcal{O}$(1) TeV& \begin{tabular}{@{}c@{}}$D_i \rightarrow e_i^+e_i^- + d_i +\nu_i +E_i$, \\ $d_i\bar{d_i}+d_i+e_i^+e_i^-$ \end{tabular} \\\hline\hline \end{tabular}
\caption{Brief descriptions of the physical scalars and fermions required for the calculations. Here $n_1$ and $n_2$ are the dark matter candidates. The $E^0$ decays only to neutral fields leading to a missing energy signal at the LHC.}
\label{table:particle}\end{table}


\section{Neutrino masses} \label{neutrino mass}

Here we summarize the neutrino mass generation mechanism and the numerical results in our model,
and refer the reader to \cite{Dutta:2018qei} for more details.

\begin{figure}[h]
\centering
\includegraphics[height=3.5cm,width=7cm]{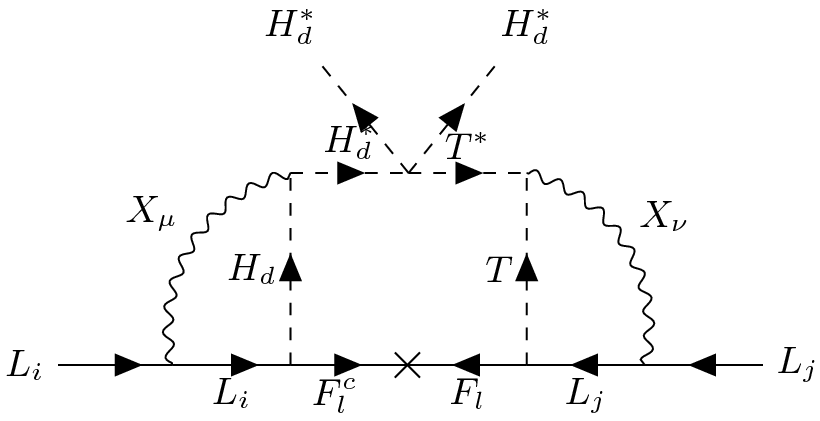}
\caption{\label{fig:neutrino:int} The three loop diagram in the interaction basis responsible for the Majorana neutrino masses.}
\end{figure}

The tree-level Lagrangian in our model contains no mass terms for neutrinos, so one must rely on radiative mass generation (i.e. through loops). Normally, the particles running in such loops need to be made heavy to achieve the vanishingly small neutrino masses, but due to the particle content and the associated $Z_2$ symmetry, the Majorana neutrino mass in our model cannot be generated below the three-loop level. 
The dimension-5 effective Majorana neutrino mass operator $L_i L_j H_d^* H_d^* /\mathcal{M}$, where $\mathcal{M}$ is some effective mass scale,  can be realized at the three-loop level as shown in Fig.~\ref{fig:neutrino:int}. The new heavy gauge bosons $X_1$ and $X_2$ play an important role in making the neutrino mass values small, and the mass matrix also gets suppressed from the loop factor $1/(16 \pi)^3$. Due to all these suppressions, the new physics associated with the tiny neutrino masses can be kept at the TeV scale.

Based on the benchmark points determined in \cite{Dutta:2018qei}, the following parameter point can generate the correct neutrino observables
\begin{eqnarray} \label{parameter} 
m_{H_1} &= & 5 ~  \textrm{TeV}, m_{H_2} =   500~ \textrm{GeV},   \ \ m_X = 5 ~ \textrm{TeV}, \ \  m_{h_2} = 268~ \textrm{GeV},\nonumber \\
  \ \ m_{T_1} &=& 500 ~ \textrm{GeV} ~,~\,  \notag  \text{tan}\beta  =50,\ \ v_1/\sqrt{2} = 7 \; \text{GeV}, \ \ \theta = 0.005\degree, \\
   m_f &=& (110,120,130)~ \textrm{GeV}, \ \ m_{E^0} = (105,110,115)~ \textrm{GeV}~.~\,  
\end{eqnarray} 

Renormalization group evolution suggest that the gauge coupling $g_2'$ is of similar strength to the SM couplings, and we fixed it to be 0.35. Note that in \cite{Dutta:2018qei}, the value of tan$\beta$ was set to be 2 in exchange for a smaller value of $y_1\times y_2 \sim 10^{-3}-10^{-4}$ (the parametric dependence of the neutrino mass matrix is $y_1\times y_2 \sin^2 \beta $). Since we will later need large Yukawa couplings to explain the magnetic moments, we have altered the original benchmark point while maintaining the neutrino masses. For the remainder of the paper we assume parameters in the neighborhood of this benchmark point.

The particle spectrum is consistent with the LHC as shown in \cite{Dutta:2018qei}. Of particular importance is that the new leptons can be abundantly produced at the LHC, but all of their decays include neutral states that are almost degenerate in mass making them extremely difficult to find in missing energy searches. This holds true for all fields in our model that are within the energy reach of the LHC and could, in principle, be subject to such constraints.


\section{The Muon and Electron Anomalous Magnetic Moments}	\label{sec:g-2}

\subsection{Background}

As mentioned in the Introduction, there is a 3.7~$\sigma$ discrepancy between the experimental results ~\cite{Bennett:2006fi,Tanabashi:2018oca} and the theoretical predictions~\cite{Davier:2017zfy,Blum:2018mom,Keshavarzi:2018mgv,Davier:2019can} of the anomalous magnetic moment of the muon, $a_\mu=(g_\mu-2)$. 
The discrepancy was found to be 
\begin{equation}
\label{muon} \Delta a_\mu = a_\mu^{exp}-a_\mu^{th} = (2.74\pm.73)\times 10^{-9}. 
\end{equation} 
 
The precision of the SM prediction will be improved in the future~\cite{Lehner:2019wvv}. An updated measurement is expected soon from Fermilab~\cite{Grange:2015fou, Fienberg:2019ddu} and J-PARC~\cite{Saito:2012zz}.

On the other hand, the direct experimental measurement of the electron anomalous magnetic moment, $a_e^{exp}$~\cite{Hanneke:2008tm}, had been in agreement with the SM prediction, $a_e^{th}$~\cite{Aoyama:2017uqe}, at the level of 1.7~$\sigma$ until recently, when an updated value of the fine structure constant $\alpha$ has been measured with high precision using Cesium atoms~\cite{Parker:2018vye}
\begin{equation} 
\alpha^{-1}(Cs) = 137.035999046(27) ~.~\,
\end{equation}

The result of this precise measurement of $\alpha$ leads to a 2.4~$\sigma$ discrepancy between the experiment~\cite{Hanneke:2008tm, Hanneke:2010au} and theory~\cite{Aoyama:2017uqe} values of $a_e$
\begin{equation} \label{electron} 
\Delta a_e = a_e^{exp}-a_e^{th} = (-8.7\pm3.6)\times 10^{-13}. 
\end{equation}

The simplest BSM explanations of these values, say a new mediator that couples to both electrons and muons, are expected to result in $\Delta a_e$ and $\Delta a_\mu$ of the same sign, since the new physics couplings would appear twice in each diagram. Moreover, if coupling universality is assumed then we additionally expect the corrections to scale with the lepton mass, that is $\Delta a_e/\Delta a_\mu \sim m_e^2/m_\mu^2 \sim 2.25\times 10^{-5}$. Since neither of those is true, a more complex solution is needed. Several such solutions exist in the literatures~\cite{Davoudiasl:2018fbb, Crivellin:2018qmi, Liu:2018xkx, Dutta:2018fge, Han:2018znu, Crivellin:2019mvj, Endo:2019bcj}.

In this work, we rely on the diversity of Yukawa couplings in Eq. \ref{eq:g-22}. These terms allow for a variety of new scalars and fermions to run in the loop, with couplings that are both chiral (different for left-handed and right-handed components) and flavor non-universal (different for each lepton). As we show in the next subsection, the chirality of these interactions leads certain couplings to appear only once in a given diagram thereby allowing for corrections to $a_e$ and $a_\mu$ in opposite directions, while the non-universality allows for modifying each independently of the other.


\subsection{Calculations and results}

We now proceed to present the calculations and results of the anomalous magnetic moments. We choose to work in the physical mass basis.  Upon expanding Eq.\ref{eq:g-22} in terms of the physical fields, the Lagrangian that generates the necessary one-loop diagrams can be written as
\begin{eqnarray} \label{eq:g-2lag}
-\mathcal{L}&=&\bar{e_i}[C_{S1}+C_{P1}\gamma_5]f_{1i}s_3
+\bar{e_i}[C_{S2}+C_{P2}\gamma_5]f_{2i}s_3
+\bar{e_i}[C_{S3}+C_{P3}\gamma_5]f_{1i}s^\prime_3\nonumber \\
&+&\bar{e_i}[C_{S4}+C_{P4}\gamma_5]f_{2i}s^\prime_3
+\bar{e_i}[C_{S5}+C_{P5}\gamma_5]e_i s_2
+\bar{e_i}[C_{S6}+C_{P6}\gamma_5]e_i s_{2s}\nonumber \\
&+&\bar{e_i}[C_{S7}+C_{P7}\gamma_5]e_i s^\prime_1
+\bar{e_i}[C_{S8}+C_{P8}\gamma_5]E^0_i H_1^-
+\bar{e_i}[C_{S9}+C_{P9}\gamma_5]E^0_i H_2^-\nonumber \\
&+&\bar{e_i}[C_{S10}+C_{P10}\gamma_5]\nu_i h_1^-
+\bar{e_i}[C_{S11}+C_{P11}\gamma_5]\nu_i h_2^- +{\rm H.C.}~,~\,
\end{eqnarray}
where the coefficients are given by
\begin{eqnarray} \label{coeff}
C_{S1} & = &
\frac{1}{2\sqrt{2}}(y_{8i}\sin \psi \cos \theta_f-y_{2i} \sin \psi \sin \theta_f +y_{6i}\cos \psi \cos \theta_f) ,
\nonumber\\
C_{P1} &= &  \frac{1}{2\sqrt{2}}(y_{8i}\sin \psi \cos \theta_f-y_{2i} \sin \psi \sin \theta_f +y_{6i}\cos \psi \cos \theta_f),
\nonumber\\
C_{S2}&=& \frac{1}{2\sqrt{2}}(y_{8i}\sin \psi \cos \theta_f-y_{2i} \sin \psi \sin \theta_f +y_{6i}\cos \psi \cos \theta_f) ,
\nonumber\\
C_{P2}&=&\frac{1}{2\sqrt{2}}(y_{8i}\sin \psi \cos \theta_f-y_{2i} \sin \psi \sin \theta_f +y_{6i}\cos \psi \cos \theta_f),
\nonumber\\
C_{S3}&=&\frac{1}{2\sqrt{2}}(y_{8i}\sin \psi \cos \theta_f-y_{2i} \sin \psi \sin \theta_f +y_{6i}\cos \psi \cos \theta_f),
\nonumber\\
C_{P3}&=&\frac{1}{2\sqrt{2}}(y_{8i}\sin \psi \cos \theta_f-y_{2i} \sin \psi \sin \theta_f +y_{6i}\cos \psi \cos \theta_f),
\nonumber\\
C_{S4}&=&\frac{1}{2\sqrt{2}}(y_{8i}\sin \psi \cos \theta_f-y_{2i} \sin \psi \sin \theta_f +y_{6i}\cos \psi \cos \theta_f),
\nonumber\\
C_{P4}&=&\frac{1}{2\sqrt{2}}(y_{8i}\sin \psi \cos \theta_f-y_{2i} \sin \psi \sin \theta_f +y_{6i}\cos \psi \cos \theta_f),
\nonumber\\
C_{S5}&=&\frac{1}{2\sqrt{2}} y_{8i}c_{12}=-C_{P5} ;~C_{S6}=\frac{1}{2\sqrt{2}}y_{8i}c_{13}=-C_{P6},
\nonumber\\
C_{S7}&=&\frac{i}{2\sqrt{2}}y_{8i}c^\prime_{13}=-C_{P7} ;~C_{S8}=\frac{1}{2}y_{8i}\cos \theta=-C_{P8},
\nonumber\\
C_{S9}&=&-\frac{1}{2}y_{8i}\sin \theta=-C_{P9} ;~C_{S10}=-\frac{1}{2}y_{8i}\cos \beta=-C_{P10},
\nonumber\\
C_{S11}&=&\frac{1}{2} y_{8i}\sin \beta=-C_{P11} ~.~\,
\end{eqnarray}

\begin{figure}[h]

\begin{subfigure}[b]{0.48\textwidth}
\includegraphics[width=0.6\linewidth,height=5cm]{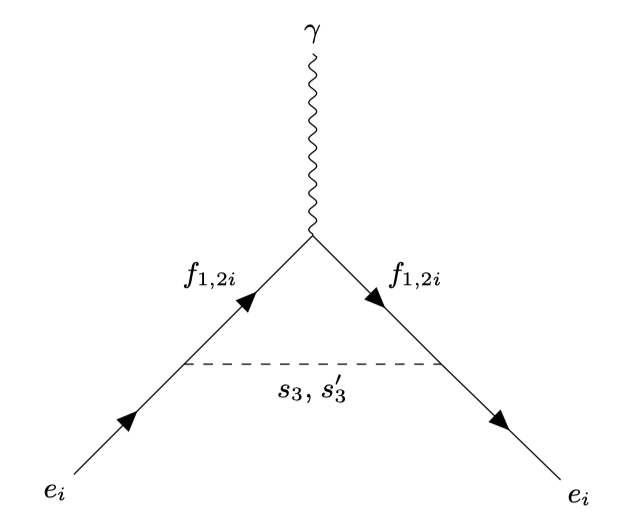}
\caption{\label{fig:1}}
\end{subfigure}
\begin{subfigure}[b]{0.48\textwidth}
\includegraphics[width=0.6\linewidth,height=5cm]{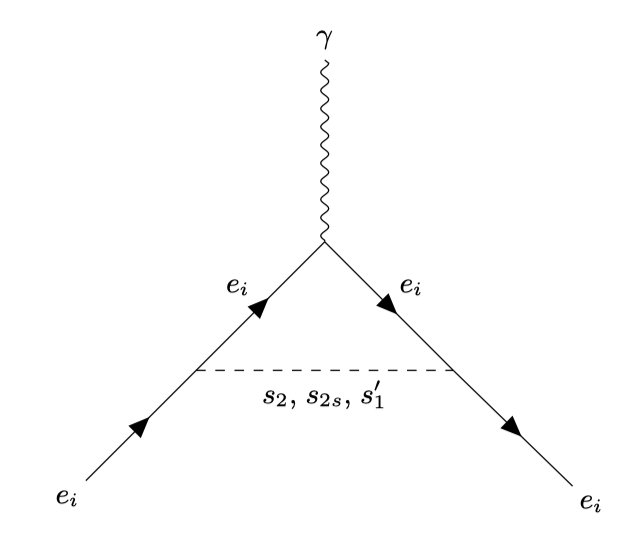}
\caption{\label{fig:2}}
\end{subfigure}
\begin{subfigure}[b]{0.48\textwidth}
\includegraphics[width=0.6\linewidth,height=5cm]{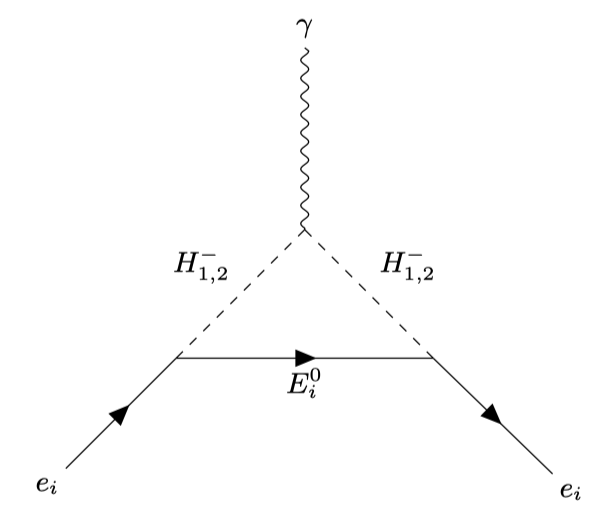}
\caption{\label{fig:3}}
\end{subfigure}
\begin{subfigure}[b]{0.48\textwidth}
\includegraphics[width=0.6\linewidth,height=5cm]{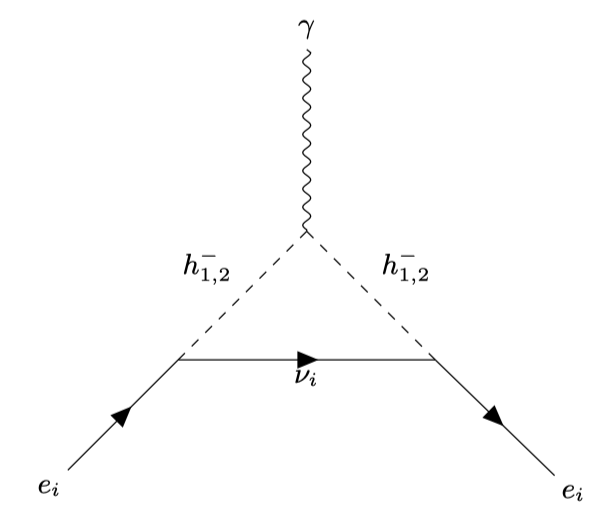}
\caption{\label{fig:4}}
\end{subfigure}
\caption{\label{fig:g-2mass} One-loop diagrams in the mass basis which contributes to the anomalous magnetic moments of the muon and electron. }
\end{figure}

The 11 different terms in Eq.~\ref{eq:g-2lag} can generate 11 different Feynman diagrams as shown in Fig.~\ref{fig:g-2mass} \footnote{We have used the package TikZ-Feynman~\cite{Ellis:2016jkw} to draw the diagrams.}. Note that the couplings $C_{S1}$-$C_{P11}$ are linear combinations of three Yukawa couplings $y_2$, $y_6$, and $y_8$. This will lead to products of two different Yukawa couplings in various diagrams.

The Feynman diagrams of Fig.~\ref{fig:g-2mass} can be broadly categorized into two category:  ones with a neutral scalar inside the loop and ones with a charged scalar. Each term in the above Lagrangian takes the general form
\begin{equation}
 -\mathcal{L} = \bar{e_i}[C_{S}+C_{P}\gamma_5]f_{i}s  ~,~\,
 \end{equation}
 where $f$ denotes the fermion and $S$ the scalar that run in the loop.
 
Following Ref.~\cite{Leveille:1977rc}, the contribution from the first type of diagrams with a neutral scalar can be written as
\begin{equation}
 \label{neutral}
 \Delta a^1_{ei}(C_S, C_P, m_{fi}, m_s) = \frac{-q_{f_i} m^2_{ei}}{8 \pi^2}\int^1_0 dx  \frac{\left[ C_S^2 \left( x^2-x^3+\frac{m_{fi}}{m_{ei}}x^2 \right)+C_P^2 \left( m_{fi} \rightarrow -m_{fi} \right) \right]} {m^2_{ei} x^2+(m^2_{fi}-m^2_{ei})x+m^2_s(1-x)},
 \end{equation}

and the contribution of the second type of diagrams with charged scalar can be written as
\begin{equation}
\label{charged} 
 \Delta a^2_{ei}(C_S, C_P, m_{fi}, m_s) = \frac{-q_h m^2_{ei}}{8 \pi^2}\int^1_0 dx  \frac{\left[ C_S^2 \left( x^2-x^3+\frac{m_{fi}}{m_{ei}}(x^2-x) \right)+C_P^2 \left( m_{fi} \rightarrow -m_{fi} \right) \right]} {m^2_{ei} x^2+(m^2_s-m^2_{ei})x+m^2_{fi}(1-x)}
  \end{equation}

In the following, we discuss the various contributions to $\Delta a_{ei}$.

\begin{itemize}
\item The first four terms of Eq.~\ref{eq:g-2lag} give four diagrams where we have the vector-like leptons $f_{1,2}$ and the new neutral scalar particles $s_{3}~\mbox{and}~s^\prime_3$ inside the loop. Fig.~\ref{fig:1} shows these diagrams and their contributions to $\Delta a_{ei}$ is given by Eq.~\ref{neutral}. From Eq.~\ref{coeff} and Eq.~\ref{neutral}, we get quadratic terms in the Yukawa couplings ($y_{2i}^2,~ y_{6i}^2~\mbox{and}~ y_{8i}^2$) as well as cross terms ($y_{2i} \times y_{8i}$, $y_{2i} \times y_{6i}$ and $y_{6i} \times y_{8i}$). The quadratic terms are proportional to $m_{ei}^2$ while the cross terms are proportional to $m_{fi}\times m_{ei}$. With a fermion mass $m_{fi} \sim 100 $ GeV, the cross terms of Yukawa couplings can lead to contributions that are both large and of opposite signs for the muon and electron cases.

\item The next three terms of Eq.~\ref{eq:g-2lag} give three more diagrams in Fig.~\ref{fig:2} with the new neutral scalars $s_{2},~s_{2s}~\mbox{and}~s^\prime_{1} $ inside the loop along with the SM muons and electrons. Together with the Hermitian conjugates, these terms become simple, 
for example, $2C_{S5}\bar{e_i}e_is_2$. Their contributions to $\Delta a_{ei}$ is given by Eq.~\ref{neutral}. All the diagrams are proportional to the SM Yukawa couplings, $y_{8i}^2$, and hence suppressed by $m_{ei}^2$. Therefore, their contribution to $\Delta a_{ei}$ is small compared to the first four diagrams. Since we do not rely on these contributions, the masses of the scalars involved are thus far not fixed. 

\item Fig.~\ref{fig:3} and \ref{fig:4} give four more diagrams arising from the last four terms in Eq.~\ref{eq:g-2lag}. All four diagrams have charged scalars $(H^\pm_{1,2}~ \mbox{and}~h^\pm_{1,2})$  and neutral fermions such as $E^0_i~\mbox{and}~\nu_i$ inside the loop. Their contributions to $\Delta a_{ei}$ are given in Eq.~\ref{charged}. These diagrams are also suppressed compared to the diagrams with cross terms due to the SM Yukawa factors $y_{8i}^2$. Additionally, since these particles also enter into the three-loop diagrams~(Fig.~\ref{fig:neutrino:int}) needed for neutrino mass generation, their masses are already fixed in our model, and so they do not play important roles in the $\Delta a_{ei}$ calculations.

\item In addition to these 11 scalar loop diagrams contributing to $\Delta a_{ei}$, we do get contributions from the gauge bosons associated with the new gauge group $SU(2)_N$. The lower limit on the new gauge boson masses is $\sim$ 3.6 TeV~\cite{Sirunyan:2018xlo, Aaboud:2017yvp, Dutta:2018qei} assuming the gauge coupling $g_2'$ is 0.35. These diagrams are also suppressed by the square of lepton masses. Therefore, these diagrams are small compared to the scalar diagrams of Fig.~\ref{fig:1} and their contributions can be neglected. 

\end{itemize}

For completeness we consider the contributions to $\Delta a_{ei}$ from all the scalar diagrams. The total contribution to $\Delta a_{ei}$ can be expressed  in a simple form
\begin{eqnarray} \label{g-2}
\Delta a_{ei} &=& \Delta a^1_{ei}(C_{S1}, C_{P1}, m_{f_{1i}}, m_{s_3})+\Delta a^1_{ei}(C_{S2}, C_{P2},  m_{f_{2i}}, m_{s_3})+\Delta a^1_{ei}(C_{S3}, C_{P3},  m_{f_{1i}}, m_{s^\prime_3})
 \nonumber\\
  &+&\Delta a^1_{ei}(C_{S4}, C_{P4},  m_{f_{2i}}, m_{s^\prime_3})+\Delta a^1_{ei}(C_{S5}, m_{ei}, m_{s_2})+\Delta a^1_{ei}(C_{S6}, m_{ei}, m_{s_{2s}})
 \nonumber\\
  &+&\Delta a^1_{ei}(C_{S7},m_{ei}, m_{s^\prime_1})+\Delta a^2_{ei}(C_{S8}, C_{P8}, m_{E_i^0}, m_{H_1})+\Delta a^2_{ei}(C_{S9}, C_{P9},  m_{E_i^0}, m_{H_2}) \nonumber\\ &+&\Delta a^2_{ei}(C_{S10}, C_{P10}, m_{\nu i }, m_{h1})+\Delta a^2_{ei}(C_{S11}, C_{P11}, m_{\nu i }, m_{h2})~.~\,
  \end{eqnarray}

In order to find a working parameter point, we begin by fixing the dimensionless parameters that are not yet set by neutrino masses and then vary the masses of new fields. The $y_8$'s are fixed by the SM charged lepton masses by the relation $m_{ei}= y_{8i}v_1/\sqrt{2}$, and we therefore set them to be $y_{8e}= -7\times 10^{-5}$, $y_{8\mu}= 0.015$ and $y_{8\tau}= 0.25$. Other necessary coupling constants can be taken as follows:~$ y_{2e} =  0.80,~~ y_{6e} = 2.50,~~ y_{2\mu} = 0.50~~  \mbox{and}~~ y_{6\mu} = 0.25 $.

For simplicity, we choose the mixing angles $\psi$, $\psi^\prime$  $\theta_f$, $\theta_{1,2,3}$ and $\theta^\prime_{1,2,3}$ to be $45^\degree$. 
In Table~\ref{table:g-2}, we give five different sets of values of the fermion and scalar masses that play an important role in the $\Delta a_{\mu}$ and $\Delta a_{e}$ calculations. As we discussed earlier, the dominant contributions to $\Delta a_{ei}$ are coming mainly from the diagrams of Fig.~\ref{fig:1}, in particular from the terms with the product of two different Yukawa couplings. This Yukawa structure allows for the sought after violation of the scaling dependance $\Delta a_e/\Delta a_\mu \propto m_e^2/m_\mu^2$, leading instead to a ratio of  $m_e/m_\mu$ and without a constraint on the sign. The particles necessary to produce this dominant contributions are $f_{1,2}$, $s_{3}$ and $s^\prime_3$. This is true for any value of the parameters in Eq.~\ref{parameter}. 

\begin{table}[h]
\centering
\begin{tabular}{ |c|c|c|c|c|c| }
\hline\hline
Benchmark Point& $m_{f1}$(GeV)& $m_{f2}$(GeV)& $m_{s3}$(GeV)& $m_{s^\prime_3} $ (GeV)\\\hline
BP1&120&121&350&1985\\\hline
BP2&120&135&350&1121\\ \hline
BP3&120&102&350&1578\\\hline
BP4&120&118&350&570\\\hline
BP5&120&145&350&2150\\\hline\hline
\end{tabular}
\caption{ \label{table:g-2} Five Benchmark points are shown which can account for the discrepancy in the anomalous magnetic moments of 
the muon and electron as well as be consistent with neutrino mass calculations.}
\end{table}

In order to get of sense of how readily the model fits the observations we perform a random scan over some of the parameters going into the $g-2$ calculation. We limit the scan to a subset of four parameters for tractability. The diagrams in Fig.~\ref{fig:g-2mass} suggest that the dependence on $m_{f_1}$ and $m_{s_3}$ is similar to that on $m_{f_2}$ and $m_{s^\prime_3}$ respectively. We, therefore, fix the former and scan over the latter. We similarly choose to fix $y_{2e,\mu}$ and vary $y_{6e,\mu}$.

\begin{figure}[h]
\centering
\includegraphics[height=8cm,width=14cm]{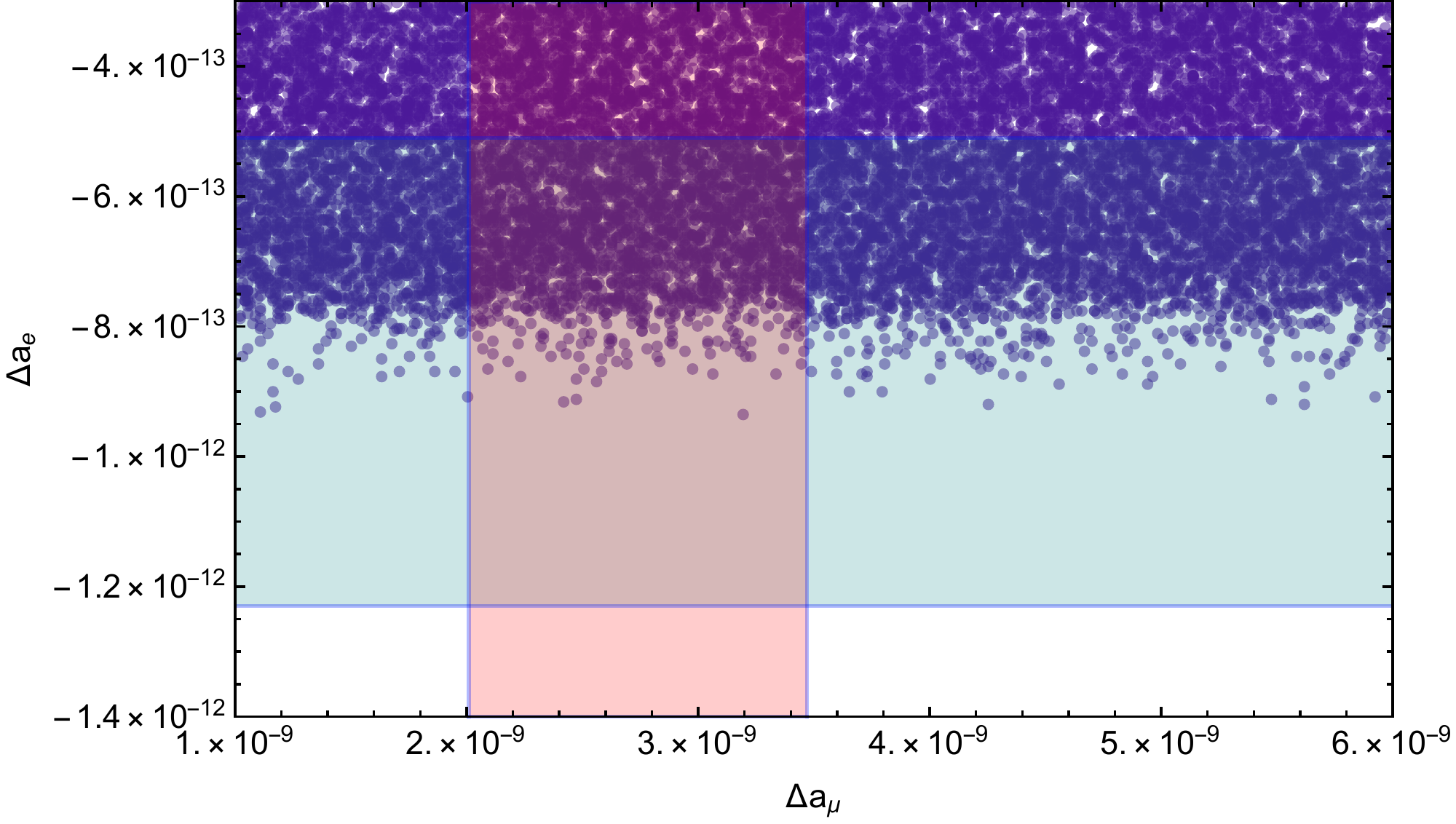}
\caption{\label{fig:plot} A scatter plot in the $\Delta a_\mu - \Delta a_e$ plane of about 16,800 points from a scan of 100,000 parameter points randomly selected from the range in Table \ref{table:scan} The bands represent the 1 $\sigma$ deviation  given by $\Delta a_\mu = (2.74\pm.73)\times 10^{-9}$ and $\Delta a_e = (-8.7\pm3.6)\times 10^{-13}$. The number of points that fell into the intersection of the two bands is 2,773.}
\end{figure}

We sample 100,000 points at random from the range shown in Table~\ref{table:scan} with $y_{2e}$ = 0.9, $y_{2\mu}$ = 0.5,$m_{f_1}$= 120 GeV and $m_{s_3}$ = 350 GeV. In Fig.~\ref{fig:plot} we show the results as a scatter plot in the $\Delta a_\mu-\Delta a_e$ plane along with the 1 $\sigma$ bands of Eqs. \ref{muon} and \ref{electron} ($\Delta a_\mu = (2.74\pm.73)\times 10^{-9}$ and $\Delta a_e = (-8.7\pm3.6)\times 10^{-13}$). About 2800 points fell into the intersection of the two bands. We can see that while a wide range of $\Delta a_\mu$ can be achieved, the values of $\Delta a_e$ mostly lie on the upper end of the band.

In experimenting with other scanning schemes we find that, indeed, allowing for the fixed parameters to vary leads to no significant expansion of the $\Delta a_e$ reach. We also find that there is very strong dependence on the value of $y_{6e}$. Namely, expanding the lower scan limit of $y_{6e}$ down to 0.01 leads to a dramatic drop in the density of viable points.

\begin{table}[h]
\centering
\begin{tabular}{ |c|c| }
\hline\hline
Parameter & Range\\\hline
$y_{6e}$&0.6-3.0\\\hline
$y_{6\mu}$&0.001-2.0\\ \hline
$m_{f_2}$&100-150 GeV\\\hline
$m_{s^\prime_3}$&300-2500 GeV\\\hline \hline
\end{tabular}
\caption{ \label{table:scan} The scan range used to generate Fig. \ref{fig:plot}. The fixed parameters are: $y_{2e}$ = 0.9, $y_{2\mu}$ = 0.5, $m_{f_1}$= 120 GeV and $m_{s_3}$ = 350 GeV.}
\end{table}

\section{Anita} \label{Anita}
\subsection{Background}

ANITA is an Antarctic balloon experiment that looks for ultra high energy cosmic rays by detecting the associated geosynchrotron emissions  using a series of radio antennas \cite{Gorham:2008dv}. The flight altitude of over 30 km leads to a $10^6\;\text{km}^2$  coverage of the Antarctic ice which compensates for the limited operation time compared to IceCube and the Pierre Auger Observatory, and offers sensitivity to a complementary range of energies and phenomena. In particular, the polarity of the geosynchroton radiation is correlated with the Earth's magnetic field, allowing ANITA to discriminate between showers emerging directly from the Earth and down going showers that are reflected off the ice. During the first and third flights of ANITA, 2 out of the 36 detected events stood out.

During its first flight, ANITA-I~\cite{Gorham:2016zah}, an event with an energy of 0.6$\pm$0.4 EeV was detected at a zenith angle of $z=117.4^\degree\pm 0.3^\degree$ with a non-inverted polarity, suggesting that the event emerged from the Earth rather than reflected off the ice. More recently, a similar event, with non-inverted polarity, has been found in the ANITA-III data ~\cite{Gorham:2018ydl}, with an energy of 0.56$^{+0.30}_{-0.20}$ EeV and a zenith angle of $z= 125^\degree \pm 0.3^\degree$. At such high energies the survivability of neutrinos passing through such a long arc length within the Earth is very low, and the isotropic neutrino flux required to explain the events with a $\nu_\tau$ that up-scatters to a $\tau$ has been found to be at least 2 \cite{Romero-Wolf:2018zxt}, and possibly 6 \cite{Fox:2018syq} orders of magnitude larger than the limits set by the Pierre Auger Observatory and IceCube. The $\tau$ emergence probability has also been studied in \cite{Alvarez-Muniz:2017mpk}

One approach around this is to replace the $\tau$ with an intermediary field with a long lifetime and large survivability in matter which decays to a $\tau$ upon or right before emergence. In our case, we rely on the new neutral field $\bar{E}^0$ which is produced via the scattering of a $\nu_\tau$ off of the Earth matter. The new heavy field can then propagate through the Earth, emerge at the South Pole, and generate a $\tau$ as part of its decay products, which can then decay hadronicaly leading to an observable signal at ANITA.  A similar approach has been considered in \cite{Huang:2018als, Dudas:2018npp, Connolly:2018ewv, Collins:2018jpg, Chauhan:2018lnq, Dev:2019ekc}. 

There are, however, several caveats to the story. It has been shown that solutions that rely on a neutrino flux still exceed the astrophysical bounds despite the higher rate of $\tau$ emergence compared to the SM only scenario (see, for example, \cite{Cline:2019snp}). This is true for an isotropic neutrino flux or a long-lasting point source, so one needs to postulate a rather exotic astrophysical source of neutrinos to fully explain the ANITA observation. Explanations that rely on a source other than a neutrino flux, say dark matter, do not suffer from this drawback \cite{Anchordoqui:2018ucj, Yin:2018yjn, Heurtier:2019git, Heurtier:2019rkz, Hooper:2019ytr, Cline:2019snp, Esteban:2019hcm, Heurtier:2019rkz, Borah:2019ciw}, although see the discussion in Ref. \cite{Chipman:2019vjm}. A summary of feasibility of various approaches can be found in the conference proceedings by Ref. \cite{Anchordoqui:2019utb}.

A second caveat is that any BSM explanation of the ANITA events will have to confront the lack of similar events in IceCube \cite{Fox:2018syq}. Depending on the mechanism utilized, the dynamics could conspire to produce a signal in one experiment but not the other. Intriguingly, building upon a proposal by Ref. \cite{Kistler:2016ask}, Ref. \cite{Fox:2018syq} shows that a small reported tension between the northern track and the full sky spectra measured by IceCube could be alleviated if some of the up-going muon events are interpreted as misidentified tau events, and they identify three such events by calculating the probability of emergence as a function of energy and angle. If true, these events would have higher energy than reported by IceCube, thus adding support to the ANITA observations. This interpretation, however, does not improve the viability of a neutrino flux explanation at either experiment.

A final caveat is that non-BSM explanations of the ANITA observations are still possible. Very recently it was proposed that an additional electromagnetic component can be emitted during the shower's transit across the ice surface with a non-inverted polarity \cite{deVries:2019gzs}. Another recent paper proposed that some ice features can lead to reflection without an inversion of polarity \cite{Shoemaker:2019xlt}. Two earlier proposals that were not able to account for the anomaly were transition radiation through the air \cite{Motloch:2016yic} and a more accurate treatment of the reflections of the electromagnetic waves \cite{Dasgupta:2018dzp}.

With all of that mind, we content ourselves with an estimate of the emergence fraction of taus in our model, using the SM fraction from \cite{Fox:2018syq} as a reference point.


\subsection{Calculations and results}

\begin{figure}[h]

\begin{subfigure}[b]{0.48\textwidth}
\includegraphics[width=0.6\linewidth,height=4.5cm]{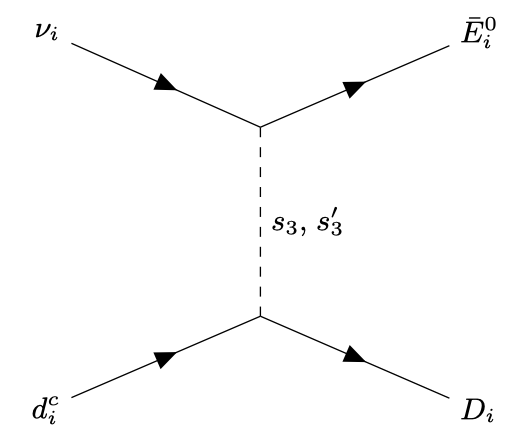}
\caption{\label{fig:ANITAcross}}
\end{subfigure}
\begin{subfigure}[b]{0.48\textwidth}
\includegraphics[width=0.7\linewidth,height=5cm]{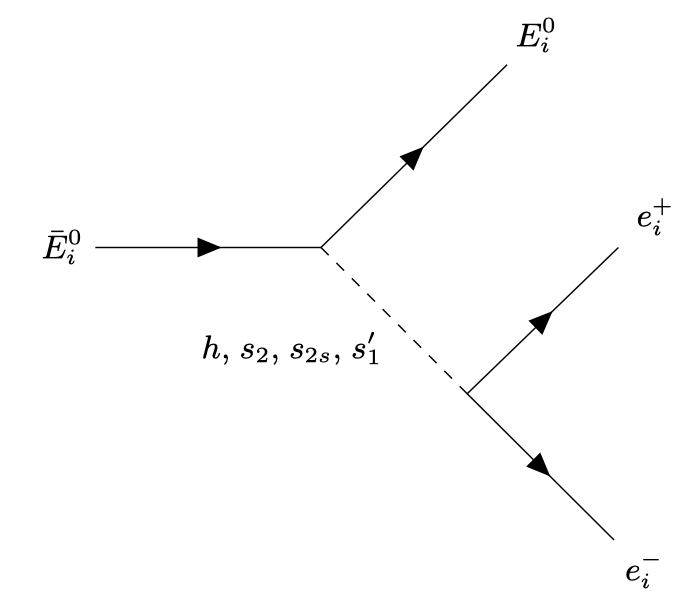}
\caption{\label{fig:ANITAdecay}}
\end{subfigure}

\caption{Feynman diagrams in mass basis for (a) neutrino-nucleon interaction and (b) the decay of the neutral  vector-like lepton to tau final state.}
\end{figure}

To enhance the rate of $\tau$ emergence, we identify a particle that can be produced from neutrino scattering off of nuclei, survive passage through Earth along a  chord length of about $l\sim$5740-7210 km corresponding to the angles of two ANITA events, and decay to a $\tau$ lepton. The neutral vector-like lepton $\bar{E}^0$ from the doublet $L^\prime$ can serve such a purpose. It can be produced in the $\nu_\tau$-nucleon scattering mediated by the neutral scalars $s_3~\mbox{and}~s^\prime_3$ along with a heavy quark $D_i$ as shown in Fig.~\ref{fig:ANITAcross}. It can then propagate through the Earth without significant attenuation and finally decay into a $\tau$ lepton pair and another heavy neutral particle as shown in Fig.~\ref{fig:ANITAdecay}. 

We start our analysis by writing the necessary  terms of the Lagrangian in the mass basis

\begin{dmath} \label{lag:anita} -\mathcal{L} = -\frac{y_{5ij} \cos \psi}{2\sqrt{2}} \bar{d_i^c}(1-\gamma_5) D_j s_3-\frac{iy_{5ij} \cos \psi^\prime}{2\sqrt{2}} \bar{d_i^c}(1-\gamma_5) D_j s^\prime_3-\frac{y_{6ij} \cos \psi}{2\sqrt{2}} \bar{E_i^0}(1-\gamma_5) \nu_j s_3-\frac{iy_{6ij} \cos \psi^\prime}{2\sqrt{2}} \bar{E_i^0}(1-\gamma_5) \nu_j s^\prime_3+\frac{y_{6ij}c_{31}}{2\sqrt{2}}\bar{E_i^0}(1-\gamma_5) E^0_j h+\frac{y_{6ij}c_{32}}{2\sqrt{2}}\bar{E_i^0}(1-\gamma_5) E^0_j s_2+\frac{y_{6ij}c_{33}}{2\sqrt{2}}\bar{E_i^0}(1-\gamma_5) E^0_j s_{2s}+\frac{iy_{6ij}c^\prime_{33}}{2\sqrt{2}}\bar{E_i^0}(1-\gamma_5) E^0_j s^\prime_1-\frac{y_{3ij}c_{11}}{2\sqrt{2}}\bar{d_i}(1-\gamma_5) d_j h-\frac{y_{3ij}c_{12}}{2\sqrt{2}}\bar{d_i}(1-\gamma_5) d_j s_2-\frac{y_{3ij}c_{13}}{2\sqrt{2}}\bar{d_i}(1-\gamma_5) d_j s_{2s}-\frac{iy_{3ij}c^\prime_{13}}{2\sqrt{2}}\bar{d_i}(1-\gamma_5) d_j s^\prime_1+\frac{y_{8ij}c_{11}}{2\sqrt{2}}\bar{e_i}(1-\gamma_5) e_j h+\frac{y_{8ij}c_{12}}{2\sqrt{2}}\bar{e_i}(1-\gamma_5) e_j s_2+\frac{y_{8ij}c_{13}}{2\sqrt{2}}\bar{e_i}(1-\gamma_5) e_j s_{2s}+\frac{iy_{8ij}c^\prime_{13}}{2\sqrt{2}}\bar{e_i}(1-\gamma_5) e_j s^\prime_1 + H.c.
	\end{dmath}
	
	The first four terms of Eq.~\ref{lag:anita} are needed for the neutrino-nucleon cross sections and the rest of the terms are responsible for the decay of $\bar{E}^0$. The Feynman diagrams of the scattering process and decay channel in the mass basis are shown in Fig.~\ref{fig:ANITAcross} and Fig.~\ref{fig:ANITAdecay}, respectively.
	
We consider cosmic neutrinos of energies about $\mathcal{O}$(EeV) which scatter off of matter producing $\bar{E}^0$ along with hadrons. In the limit of large momentum transfer the differential cross section is given by
\begin{equation}
 \label{diffcross}
\frac{d^2\sigma}{dxdy}= \frac{3y_5^2 y^2_{6 \tau}}{256\pi ME_\nu x^2} \left( \cos^2\psi -\cos^2 \psi^\prime \right)^2\sum_i[xf_i(x,\sqrt{2ME_\nu xy})+x\bar{f}_i(x,\sqrt{2ME_\nu xy})]~,~\,
\end{equation}
where $Q^2=2ME_\nu xy$ is the momentum transfer, and $f_i(x,Q^2)$ and $\bar{f}_i(x,Q^2)$ are the parton distribution functions (pdf) of the quarks and anti-quarks, respectively. $M$ is the nucleon mass, $E_\nu$ is the incoming neutrino energy, while $x$ and $y$ are dimensionless variables defined as 
\begin{equation} 
x= \frac{Q^2}{2M\nu}~~~\mbox{and}~~~y=\frac{\nu}{E_\nu} ~,~\,
\end{equation}
where $\nu=E_\nu - E_{\bar{E}^0}$ is the energy loss in the laboratory frame. The momentum transfer is always greater than the nucleon mass.  $y$ is the fractions of the initial neutrino energy transferred to the hadrons.   The total cross section can be obtained by integrating Eq.~\ref{diffcross} over $x$ and $y$
\begin{equation} \label{cross}
    \sigma = \int_{y_{min}}^1 dy \int_{x_{min}}^1 dx \frac{d^2\sigma}{dxdy}~,~\,
\end{equation} 
where $x_{min}=Q^2_{min}/2ME_\nu$, and $y_{min}$ is determined by the relation $Q^2=2ME_\nu x y$ as $y_{min}=x_{min}/x$.  The total cross section in Eq.~\ref{cross} can be calculated numerically with the given input parameters of the model and the pdfs of the quarks and antiquarks. We use the CTEQ5 parton distribution~\cite{Lai:1999wy} in our numerical calculations. We also cross checked our calculations against the analytical form of the pdfs~\cite{Berger:2007vf} and found agreement. The cross sections corresponding to three different benchmark points are shown in Table~\ref{table:ANITA}. The coupling constants are taken to be $y_{5} = 0.23$ and $y_{6\tau} = 0.17$. The cross section in Eq.~\ref{diffcross} is sensitive to the mixing angles $\psi$ and $\psi^\prime$. These angles also appear in the $g-2$ calculations but Eq.~\ref{g-2} is not sensitive enough to these angles to change the numerical results. Therefore we can vary them to get different cross sections from Eq.~\ref{cross}. We choose three different values of $\psi=$ $55^\degree$, $60^\degree$ and $65^\degree$, and $\psi^\prime=5^\degree$ corresponding to the three Benchmark Points in Table~\ref{table:ANITA}.

The $\bar{E}^0$ produced in the collision propagates through the Earth and eventually decays dominantly  into $\tau$ lepton pair and $E^0$. The three-body decay channel is mediated by the neutral scalars $h, s_2, s_{2s}$ and $s_1^\prime$. The  decay width is given by
\begin{dmath} \label{decaywidth} \Gamma_{\bar{E}^0 \rightarrow E^0 \tau \tau} =  \frac{y^2_{6 \tau} y^2_{8 \tau}}{128\pi^3} \left( \frac{c_{31} c_{11}}{m^2_{h}} + \frac{c_{32} c_{12}}{m^2_{s_2}}+ \frac{c_{33} c_{13}}{m^2_{s_{2s}}}- \frac{c_{33}^{\prime } c_{13}^{\prime }}{m^2_{s^\prime_1}}\right)^2 \times \int_{E_{E^0}^{min}}^{E_{E^0}^{max}} dE_{E^0} \int_{E_{\tau}^{min}}^{E_{\tau}^{max}} dE_{\tau} f(E_{E^0}) ~,~\, \end{dmath} 
where \begin{equation} f(E_{E^0})= \frac{E_{E^0}}{2} \left[m_{\bar{E}^0}^2-2E_{E^0}m_{\bar{E}^0}+m_{E^0}^2 -2m_{\tau}^2 \right]~,~\,  
\end{equation}
\begin{equation} E_{\tau}^{max,min} = \frac{1}{2}\left( m_{\bar{E}^0} - E_{E^0} \right) \pm \frac{1}{2}\sqrt{\left( E_{E^0}^2-m_{E^0}^2 \right)\left( 1-\frac{4m_{\tau}^2}{m_{\bar{E}^0}^2-2E_{E^0}m_{\bar{E}^0}+m_{E^0}^2} \right)} ~,~\,\nonumber \end{equation} \begin{equation} m_{E^0} \le E_{E^0}\le \frac{m_{\bar{E}^0}^2+m_{E^0}^2-4m_{\tau}^2}{2m_{\bar{E}^0}}~.~\, \end{equation}

Eq.~\ref{decaywidth} gives the sum over all four contributions for  $h, s_2, s_{2s}$ and $s_1^\prime$. The same decay channel is open for all charged leptons and down type quarks in the SM. We choose $m_{\bar{E}^0}$ and $m_{{E}^0}$ such that the bottom quark final state is kinematically forbidden. The decays to muons and strange quarks are  suppressed by a factor 100 due to the SM Yukawas. The lifetime of $\bar{E}^0$ in the rest frame is $\tau_{\bar{E}^0} = 1/\Gamma$, where $\Gamma$ is the total decay width of $\bar{E}^0$. Note that the masses of the new scalar fields are set to be $\mathcal{O}$(500) GeV, making their contributions to the decay width sub-leading compared to that of the SM Higgs. We show the lifetime for a few benchmark points in Table~\ref{table:ANITA}, taking the coupling constants to be $y_{8\tau} = 0.25$ and $y_{6\tau} = 0.17$.
 
 Let us now define the survival probability of emergence, $\epsilon_{BSM}$. We follow closely the treatment in Ref. \cite{Collins:2018jpg}. We assume that a fraction of neutrinos has survived the SM interactions after propagating a distance of $l_1$ km. We take the SM interaction length to be $l_{SM}= 275$ km~\cite{Fox:2018syq}. These surviving neutrinos can produce $\bar{E}^0$ as a result of the collision with nucleons inside the Earth. The interaction length of the $\nu_\tau$-nucleon scattering process can be defined as $l_{BSM}=1/(\sigma N_A \rho)$, where $\sigma$ is the scattering cross section, $N_A$ is the Avogadro number, $6.022\times 10^{23}$ and $\rho$ is the density of target material. For simplicity, we assume the Earth to have uniform density and take the value to be $\rho = 5 $ gm/cm$^3$. The $\bar{E}^0$ then travels a distance of $(l-l_{1})$ km, where $l=$6475 km is the average chord length along the Earth corresponding to the two ANITA events, and finally decays below an altitude of about 10 km, otherwise the air shower will not get a chance to fully develop \cite{Collins:2018jpg}. The decay length in the Earth frame is $l_{decay} = \gamma c \tau_{\bar{E}^0}$. We can now define the survival probability as
\begin{eqnarray} 
\epsilon_{BSM} &=& \int_0^l dl_1 \int_{l-l_1}^{l-l_1+d} dl_2 \int_{0}^{l-l_1-l_2+d} dl_3 \left[ \left(  \frac{e^{-\frac{l_3}{l_{decay,\tau}}}}{l_{decay,\tau}} \text{Br}(\tau \rightarrow \text{hadrons}) \right) \right. \nonumber \\ 
&\times& \left. \frac{e^{-\frac{l_2}{l_{decay}}}}{l_{decay}}\frac{e^{-\frac{l_1}{l_{BSM}}}}{l_{BSM}} \left( 1- \int_0^{l_1} {dl_4} \frac{e^{-\frac{l_4}{l_{SM}}}}{l_{SM}}\right)  \right]~,~\,
 \end{eqnarray}
where from right to left, the first expression is the fraction of $\nu_\tau$s that survive passage through matter until the interaction point, $l_1$, where $\bar{E}^0$ is produced. The second factor is the fraction of neutrinos converted to $\bar{E}^0$ anywhere inside the Earth between the emergence point $l_1=0$ and the entire chord length $l_1 = l$. The third factor is the fraction of those $\bar{E}^0$ that decay after a sufficient distance $l_2$ to emerge from the Earth, $l_2 = l-l_1$ and a further distance, $d$, beyond which it would exceed the 10 km altitude. We take $d=10/\text{sin}30\degree=20$ km. The fourth expression is the fraction of $\tau$ that decays into hadrons within the distance $d=$ 20 km.	 The decay length of $\tau$ is $l_{decay,\tau}=\gamma c \tau_\tau$, where the lifetime of $\tau$ is $\tau_\tau=$ 0.3$\times 10^{-12}$ s  and the branching ratio of $\tau$ to hadrons is $\text{Br}(\tau \rightarrow \text{hadrons})=$ 0.65 ~\cite{Tanabashi:2018oca}.

We dial our parameters such that both $l_{BSM}$ and $l_{decay}$ are comparable to the chord length. The corresponding scattering cross section, lifetime and the estimated survival probability  are shown in Table~\ref{table:ANITA}. The expression above could yield a higher value if we choose a much shorter $l_{BSM}$ and longer $l_{decay}$, but then one must account for the probability of $\nu$ regeneration. The interaction length of regeneration is similar to $l_{BSM}$ at such energies, and once accounted for the gains from reducing $l_{BSM}$ are countered and one gets similar values of $\epsilon_{BSM}$. For comparison, the SM survivability found by \cite{Fox:2018syq} are $4.4\times10^{-7}$ and $3.2\times10^{-8}$ for the ANITA-I and ANITA-III events respectively at $E_\nu  = 1$ EeV. Note that the results  in Table~\ref{table:ANITA} are far less sensitive to the chord length compared to the survivability in the SM.

\begin{table}[h]

\centering
\begin{tabular}{ |c|c|c|c|c|c|c|c|c|c|c|c|c|c|c|c| }
\hline\hline
Point& $m_{s_3}$& $m_{s^\prime_3}$&$m_{h}$& $m_{s_2}$& $m_{s_{2s}}$& $m_{s^\prime_1}$& $m_{D}$& $m_{E^0}$ &$m_{\bar{E}^0}$&$\sigma$(pb)&$\tau_{\bar{E}^0}$(ns)&$l_{BSM}$(km)&$l_{decay}$(km)&$\epsilon_{BSM}$\\\hline
BP1&500&1200&125&500&500&500&1000& 105&110&708&2&4705&6387&$2\times10^{-5}$\\\hline
BP2&500&600&125&400&400&500&1000&110&115&887&2&3757&6368&$3\times10^{-5}$\\ \hline
BP3&600&800&125&500&600&800&1000&115&120&1066&2&3126&6350&$4\times10^{-5}$\\\hline\hline

\end{tabular}
\caption{\label{table:ANITA} Three benchmark points of parameter space are shown. The parameter space can explain the ANITA events and consistent with neutrino mass and $g-2$ calculations. The masses are in GeV unit. }
\end{table}


\section{Conclusion} \label{Conclusion}

We were able to explain the observed values of the anomalous magnetic moments of the muon and electron within the framework of an $E_6$ inspired GUT model. In addition to the theoretical appeal of fitting into a unification picture, the model has been shown in an earlier work to generate neutrino masses radiatively at the 3-loop level, thereby allowing for the new physics mass scale to be within the reach of the current and proposed collider experiments. The flavor structure of the model, which allowed the neutrino fit, played a crucial role to explain the anomalous magnetic moments of the muon and electron. In this model, the ratio of $\Delta a_\mu$ and $\Delta a_e$ is  proportional to $m_\mu/m_e$ with a relative sign difference between them  as required by the experiments. We have also studied the model's contribution to the anomalous events observed by the ANITA experiments using the mass scale of new physics set by the neutrino fitting and found a modest improvement over the SM predictions. 

We stress that in explaining all the aforementioned observations we have introduced various new, SM charged particles in the 100 GeV to 1 TeV range. The primary reason these have not been ruled out is that they are too mass degenerate with one of their decay products to be covered by existing LHC searches. This adds to the list of motivations of why the efforts put in closing this mass gap in collider searches are worth pursuing.

 \textbf{Acknowledgments}

MA, BD, and SG are supported in part by the DOE Grant No. DE-SC0010813. 
TL is supported in part by the Projects 11647601 and 11875062 supported 
by the National Natural Science Foundation of China, and by 
the Key Research Program of Frontier Science, CAS.


\bibliographystyle{apsrev4-1.bst}
\bibliography{gm}

\end{document}